\documentclass[journal]{IEEEtran}

\usepackage{graphicx}
\usepackage{ulem}
\usepackage{subfigure} 
\usepackage{epsfig}
\usepackage{color}

\begin{document}
%
% paper title
\title{The LWA1 Radio Telescope %\\ {\small D R A F T -- 120419b}
      }
%
% author names and IEEE memberships
% note positions of commas and nonbreaking spaces ( ~ ) LaTeX will not break
% a structure at a ~ so this keeps an author's name from being broken across
% two lines.
% use \thanks{} to gain access to the first footnote area
% a separate \thanks must be used for each paragraph as LaTeX2e's \thanks
% was not built to handle multiple paragraphs
\author{S.W.~Ellingson,~\IEEEmembership{Senior~Member,~IEEE}, %<-this % stops a space
        G.B.~Taylor,
        J.~Craig,~\IEEEmembership{Member,~IEEE}, %<-this % stops a space
        J.~Hartman,
        J.~Dowell, %<-this % stops a space
        C.N.~Wolfe,~\IEEEmembership{Student~Member,~IEEE}, %<-this % stops a space
        T.E.~Clarke, %
        B.C.~Hicks,~\IEEEmembership{Member,~IEEE},  %
        N.E.~Kassim, %
        P.S.~Ray, %
        L.~J~Rickard, %<-this % stops a space
        F.K.~Schinzel  %{\it (and others who opt in)
        and K.W.~Weiler  
%}% <-this % stops a space
\thanks{
S.W.\ Ellingson and C.N.\ Wolfe are with the Bradley Dept.\ of Electrical \& Computer Engineering, Virginia Polytechnic Institute \& State University, Blacksburg, VA 24061 USA (e-mail: ellingson@vt.edu).  
G.B.\ Taylor, J.\ Craig, J.\ Dowell, L.~J Rickard, and F.K.\ Schinzel are with the Dept.\ of Physics \& Astronomy, University of New Mexico, Albuquerque, NM 87131 USA (e-mail: gbtaylor@unm.edu).
J.\ Hartman is with NASA Jet Propulsion Laboratory, Pasadena, CA 91109.
T.E.\ Clarke, B.C.\ Hicks, N.E.\ Kassim, and P.S.\ Ray are with the U.S. Naval Research Laboratory, Washington, DC 20375 (e-mail: namir.kassim@nrl.navy.mil).
K.W.\ Weiler is with Computational Physics, Inc., Springfield, VA 22151.
}
}
% note the % following the last \IEEEmembership and also the first \thanks - 
% these prevent an unwanted space from occurring between the last author name
% and the end of the author line. i.e., if you had this:
% 
% \author{....lastname \thanks{...} \thanks{...} }
%                     ^------------^------------^----Do not want these spaces!
%
% a space would be appended to the last name and could cause every name on that
% line to be shifted left slightly. This is one of those "LaTeX things". For
% instance, "A\textbf{} \textbf{}B" will typeset as "A B" not "AB". If you want
% "AB" then you have to do: "A\textbf{}\textbf{}B"
% \thanks is no different in this regard, so shield the last } of each \thanks
% that ends a line with a % and do not let a space in before the next \thanks.
% Spaces after \IEEEmembership other than the last one are OK (and needed) as
% you are supposed to have spaces between the names. For what it is worth,
% this is a minor point as most people would not even notice if the said evil
% space somehow managed to creep in.
%
% The paper headers
%\markboth{IEEE TRANSACTIONS ON ANTENNAS AND PROPAGATION,~Vol.~x, No.~x,~Month~YYYY}{Shell \MakeLowercase{\textit{et al.}}: The LWA1 Radio Telescope}
\markboth{Accepted by IEEE Transactions on Antennas and Propagation. \copyright 2013 IEEE.}{Shell \MakeLowercase{\textit{et al.}}: The LWA1 Radio Telescope}
% The only time the second header will appear is for the odd numbered pages
% after the title page when using the twoside option.
% 
% *** Note that you probably will NOT want to include the author's name in **% *** the headers of peer review papers.                                   ***

% If you want to put a publisher's ID mark on the page
% (can leave text blank if you just want to see how the
% text height on the first page will be reduced by IEEE)
%\pubid{0000--0000/00\$00.00~\copyright~2009 IEEE}

% use only for invited papers
%\specialpapernotice{(Invited Paper)}

% make the title area
\maketitle

\begin{abstract}
LWA1 is a new radio telescope operating in the frequency range 10--88 MHz, located in central New Mexico.  The telescope consists of 258 pairs of dipole-type antennas whose outputs are individually digitized and formed into beams. Simultaneously, signals from all dipoles can be recorded using one of the instrument's ``all dipoles'' modes, facilitating all-sky imaging.  Notable features of the instrument include high intrinsic sensitivity ($\approx6$~kJy zenith system equivalent flux density), large instantaneous bandwidth (up to 78~MHz), and 4 independently-steerable beams utilizing digital ``true time delay'' beamforming.  This paper summarizes the design of LWA1 and its performance as determined in commissioning experiments.  
We describe the method currently in use for array calibration, and report on measurements of sensitivity and beamwidth.
\end{abstract}

% Note that keywords are not normally used for peerreview papers.

% For peer review papers, you can put extra information on the cover
% page as needed:
% \begin{center} \bfseries EDICS Category: 3-BBND \end{center}
%
% For peerreview papers, inserts a page break and creates the second title.
% Will be ignored for other modes.
\IEEEpeerreviewmaketitle

%========================================================
\section{\label{sIntro}Introduction}
%========================================================

\begin{figure}
\begin{center}\psfig{file=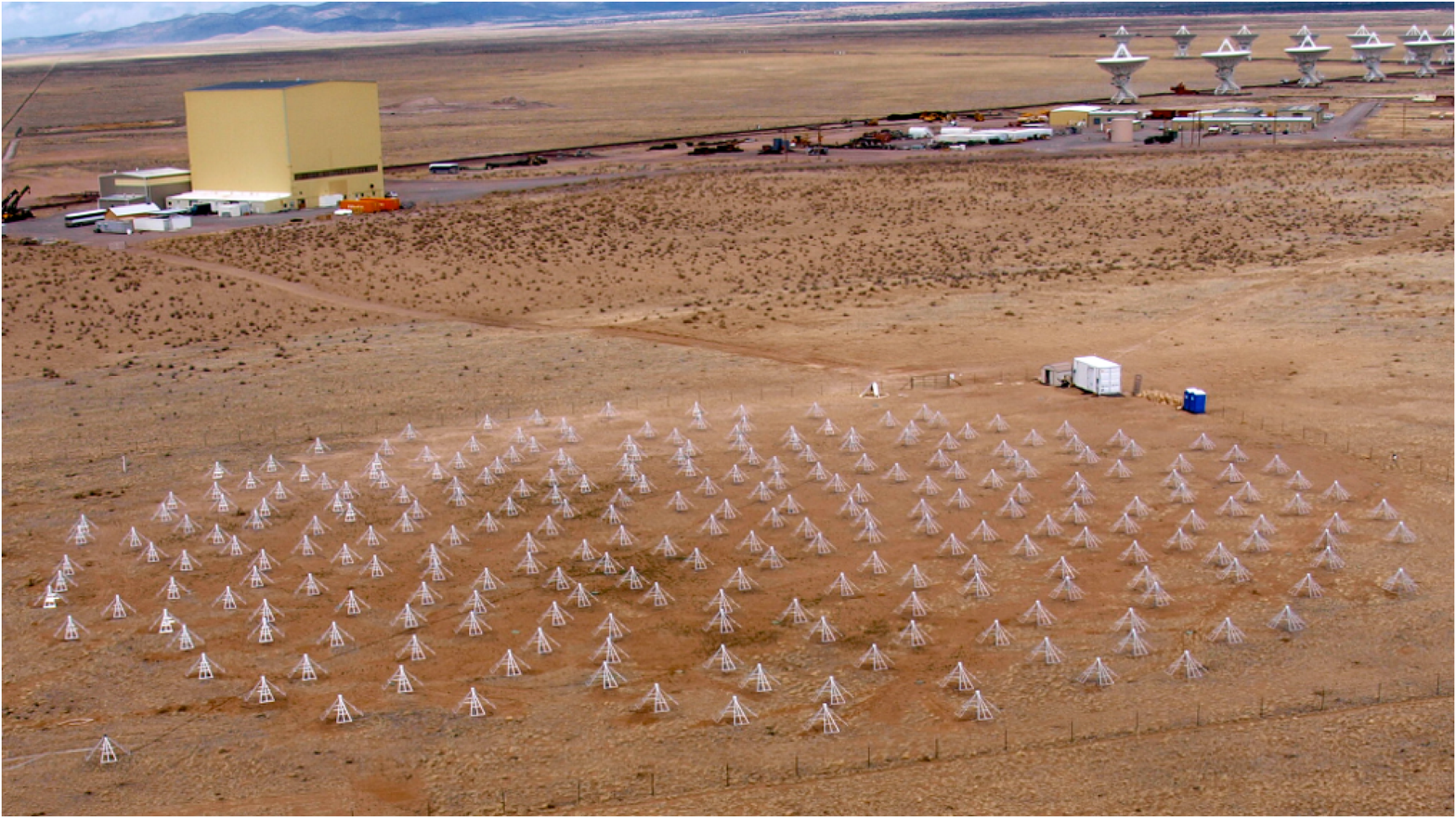,width=3.4in}\end{center} 
\caption{\label{fLWA1} LWA1.  The white cargo container beyond the station array is the electronics shelter.  Visible in the background is the center of the VLA.}
\end{figure}

LWA1 (``Long Wavelength Array Station 1''; Figure~\ref{fLWA1}) is a new radio telescope operating in the frequency range 10--88 MHz, collocated with the Very Large Array (VLA; $107.63^{\circ}$~W, $34.07^{\circ}$~N) in central New Mexico.  The telescope consists of an array of 258 pairs of dipole-type antennas whose outputs are individually digitized and formed into beams.  The principal technical characteristics of LWA1 are summarized in Table~\ref{tSpecs}.  LWA1 is so-named because it is envisioned to be the first ``station'' of a 53-station long-baseline aperture synthesis imaging array known as the Long Wavelength Array (LWA), described in \cite{PIEEE_LWA,K05}.  Although the future of the LWA is uncertain, LWA1 was completed in Fall 2011 \cite{LWAFL} and is currently operating under the U.S.\ National Science Foundation's ``University Radio Observatories'' program.
\begin{table}
\begin{center}
\begin{tabular}{|l|l|}
\hline
Beams 	                 & 4, independently-steerable \\
\hline
Polarizations            & Dual linear \\
\hline
Tunings 	         & 2 center frequencies per beam,\\
                         & independently-selectable\\
\hline
Tuning Range 	         & 24--87~MHz \\
                         & ($>$4:1 sky-noise dominated),\\
                         & 10-88 MHz usable \\
\hline
Bandwidth                & $\le16$ MHz $\times$ 2 tunings $\times$ 4 beams \\
\hline
Spectral Resolution 	 & Time-domain ``voltage'' recording; also\\
                         & real-time 1024-channel spectrometer. \\
\hline
Beam FWHM 	         & $< 3.2^{\circ}\times\left[(\mbox{74 MHz})/\nu\right]^{1.5}$ \\
                         & (upper bound independent of $Z$)\\
\hline
Beam SEFD 	         & $\approx6$~kJy at $Z=0$; depends on pointing, \\ 
                         & celestial coordinates, \& frequency; \\
                         & see Figure~\ref{fSEFD}\\
Beam Sensitivity         & $\approx8$~Jy ($5\sigma$) for 1~s, 16~MHz, $Z=0$\\
                         & (inferred from SEFD)\\
\hline
All-Dipoles Modes 	 & ``TBN'': 70~kHz from every dipole,\\
                         & continuously \\
                         & ``TBW'': 78~MHz from every dipole,\\
                         & in 61~ms ``bursts'' every 5 min \\
\hline
\end{tabular}
\end{center}
{\it Notes:} $Z$ is zenith angle. $\nu$ is frequency. 1~Jy = $10^{-26}$~W~m$^{-2}$~Hz$^{-1}$. FWHM is full-width at half-maximum.  SEFD is system equivalent flux density (see text).  Additional information available at {\tt http://lwa1.info} and {\tt http://lwa.unm.edu}.
%\begin{itemize}
%\item $Z$ is zenith angle.
%\item $\nu$ is frequency.
%\end{itemize}
\caption{\label{tSpecs}LWA1 Technical Characteristics.}
\end{table}

Contemporary radio telescopes which are also capable of operating in LWA1's 10--88~MHz frequency range include GEETEE (35--70~MHz), located in Gauribidanur, India \cite{GEETEE}; UTR-2 (5--40~MHz), located in the Ukraine \cite{UTR2}; VLA (74~MHz) \cite{VLA74}; and LOFAR (10--80~MHz), another new telescope located in the Netherlands \cite{LOFAR09,WV11}.  
%LOFAR, VLA, and LWA1 are all at least 2 orders of magnitude more sensitive than either GEETEE or UTR-2.  
LWA1 and LOFAR are both digital beamforming arrays consisting of large numbers of dipole-type antennas and comparable sensitivity; however the entire collecting area of LWA1 is contained within a single station, whereas LOFAR is an aperture synthesis array consisting of many smaller stations distributed over a large region of Northern Europe.  
See \cite{LWAFL} for a more detailed comparison of these instruments. 
%Section~\ref{sConc}. 

This paper describes LWA1 design and performance as determined in commissioning experiments.  
First, in Section~\ref{sBackground}, we provide a brief primer on radio astronomy below 88~MHz in order to provide context for subsequent discussion. 
%and briefly review past and contemporary instruments operating in this frequency regime in order to provide context.  
Section~\ref{sDesign} summarizes the design of LWA1.   
%As a very large ground-fixed array of antennas subject to significant levels of mutual coupling, calibration is not necessarily straightforward.  To this end, 
Section~\ref{sCal} describes a simple method for array calibration which we have found to be effective.  This method uses single-dipole observations of strong discrete astronomical sources correlated with observations made using an ``outrigger'' dipole located tens to hundreds of wavelengths away.   
Section~\ref{sBeam} reports the results of beamforming experiments, and compares the results to those predicted in previous LWA1 design and simulation studies \cite{PIEEE_LWA,Ellingson11}.  Conclusions are presented in Section~\ref{sConc}.

%========================================================================
\section{\label{sBackground}Background: Radio Astronomy below 88~MHz}
%========================================================================

The science applications of compact array beamforming and small-aperture imaging below 88~MHz are summarized in \cite{GWGB00} and \cite{KPJH-ASP345}; they include the study of pulsars, Jupiter, the Sun, and the Earth's ionosphere; studies of the interstellar medium including radio recombination lines and electron density variations; cosmology through observations of the redshifted 21 cm line of neutral hydrogen;  and searches for as-yet undetected phenomena including radio emission from extrasolar planets, astrophysical explosions from a variety of mechanisms, and other sources of time-variable or impulsive emission.   

Key issues for antenna and receiver design for radio astronomical instrumentation operating in this frequency range are described in \cite{Ellingson04}, which we briefly summarize here.  In this frequency range, natural external noise is dominated by the very bright Galactic synchrotron background, which generates antenna temperatures on the order of $10^3$~K to $10^5$~K, increasing with decreasing frequency.  Contributions from the cosmic microwave background ($\approx 3$~K) and other sources are also present, but are typically negligible in comparison.  Man-made noise (other than deliberate radio signals) 
%(addressed later in this paper) 
is also negligible for the rural locations at which these instruments are typically deployed.  Mechanically-steered dishes prevalent at higher frequencies are not desirable due to the very large size required to achieve acceptably narrow beamwidth.  Instead, arrays of low-gain dipole-type antennas are used to facilitate electronic or digital beamforming.\footnote{The VLA 74~MHz system uses the 27 25-m dishes of the VLA with dipole feeds, which yields very low aperture efficiency. This is tolerated as it allows reuse of existing high-frequency infrastructure.}  Individual dipole-type antennas typically have impedance bandwidth which is much less than the approximately 9:1 bandwidth implied by a tuning range of 10--88~MHz; however this is not a limitation as long as receiver noise temperature is sufficiently small that the ratio of external to internal noise is large after the antenna impedance mismatch.  A dipole-type antenna combined with a receiver having system temperature less than $\approx 500$~K is able to achieve the best possible (i.e., Galactic noise-limited) sensitivity over a large portion of the 10--88~MHz frequency range.  This is demonstrated by example in \cite{ESP07} and later in this paper.
%In other words, usable bandwidth is determined primarily by the noise figure of the receiver, and 

\begin{figure}
\begin{center}
\psfig{file=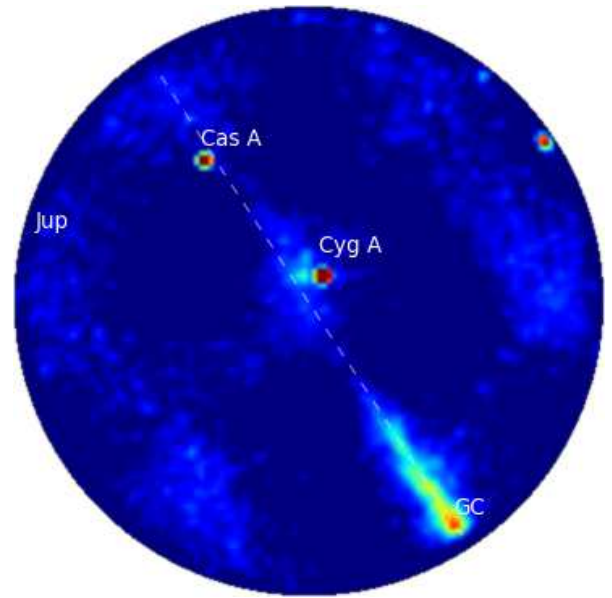,width=3.1in}
%\psfig{file=figs/120331_1625_lwatv.eps,width=3.1in}
%~\\ {\it \small [Intent is for this image to correspond to the sky around the time of the experiments discussed in later in this paper.]}
\end{center} 
\caption{\label{fPASI1} Intensity of the 74.03 MHz radio sky as measured using LWA1's PASI backend (see Section~\ref{sDesign}), 20:00 local sidereal time (LST).  The center of this display is the zenith, the perimeter is the horizon, North is up, East is left.  
The color scale ranges from $\sim2000$~K (dark blue) to $\sim250,000$~K (red).
The Galactic center (``GC'' in this display) is prominent near the bottom of the figure, and the Galactic plane extends high into the sky, as indicated by the dashed line. Bandwidth: 75~kHz, Integration time: 5~s. ``Jup'' is Jupiter, which is located just below the horizon, and the unlabeled source in the upper right is interference.  The lighter blue regions along the lower left and upper right are image noise associated with the point spread function, which has not been deconvolved from these images.}
\end{figure}

\begin{figure}
%\vspace{1in}
%{\it \small [Intent is for this image to correspond to a time when the sky is relatively boring (Galaxy is mostly ``down''.)]}
%\vspace{1in}
\begin{center}\psfig{file=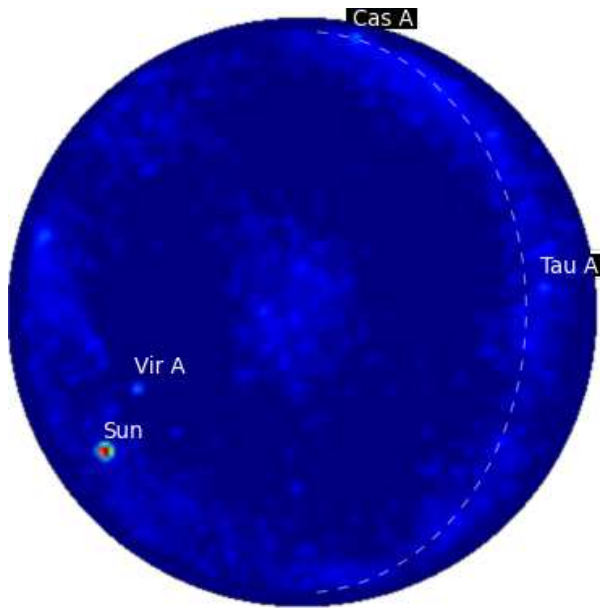,width=3.1in}\end{center} 
\caption{\label{fPASI2} Same as Figure~\ref{fPASI1}, but 14 hours later (10:00 LST).  At this time, the Galactic center is below the horizon and the Galactic plane is low on the horizon.}
\end{figure}

External-noise dominance makes the noise measured at the output of antenna-receiver channels signiﬁcantly correlated \cite{Ellingson11}.  
%This is useful for low-resolution astronomical imaging, since this correlation represents precisely the information of interest.
%as it reduces the integration time required to achieve a desired sensitivity.  In practice, however -- 
The dominance of this correlation over internal noise complicates array calibration because the strongest discrete astrophysical sources which would otherwise be suitable as ``incident plane wave'' calibrators are orders of magnitude weaker than the non-uniform Galactic noise background.  
%One alternative, outside the scope of this paper but currently being explored, is to calibrate the array by comparison of correlations between antennas to those predicted using a sky model assuming an ideally-calibrated system.  This approach is somewhat difficult due to (among other problems) the low accuracy of sky models available at these frequencies.  
The method described in Section~\ref{sCal} of this paper bypasses this problem by correlating antennas in the array with a distant outrigger antenna, which has the effect of suppressing the contribution of bright features which are distributed over large angular extent -- in particular, the Galactic background emission which is concentrated along the Galactic plane and which is brightest at the Galactic center (see Figures~\ref{fPASI1} and \ref{fPASI2}).

Separately from difficulty in calibration, external noise correlation significantly desensitizes the beams formed by the array \cite{Ellingson11}.  
%The traditional result from array signal processing theory that the signal-to-noise ratio of a beamforming array scales linearly with number of antennas, which is true when noise is dominated by amplifiers and thus not correlated between antenna-receiver chains, is not valid in this case.  
Furthermore, the extent to which a beam is desensitized is a function of sidereal time, since the Galactic noise intensity varies both spatially and diurnally, as is shown in Figures~\ref{fPASI1} and \ref{fPASI2}. This is particularly frustrating as absolute calibration of in-beam flux density is often desired, but now depends on both pointing in zenith angle ($Z$) due to the beam pattern variation as a function of $Z$, and with sidereal time due to the spatially- and diurnally-varying Galactic noise intensity arriving through sidelobes.
 
%========================================================
\section{\label{sDesign}Design}
%========================================================

%This section provides a brief review of the LWA1 design.  
%Additional details are available through the references cited.
 
\begin{figure}
\begin{center}
%\vspace{3in}
\psfig{file=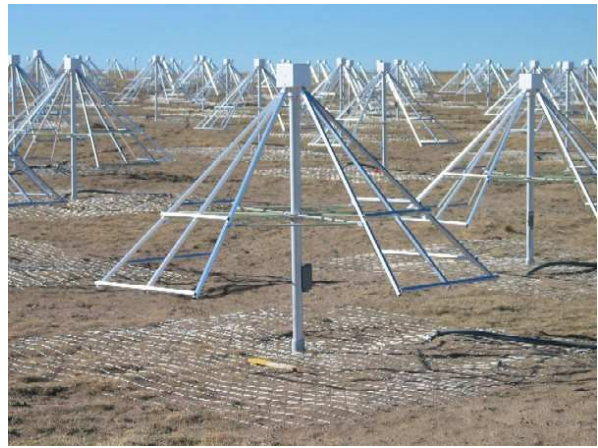,width=3.1in}
\end{center}
\caption{\label{fStand}
LWA1 antenna stands.  Front end electronics are enclosed in the white box at the feedpoint.  Signals exit through coaxial cables inside the mast.
Each stand is about 1.5~m high.
}
\end{figure}

{\it Antennas.} LWA1 antennas are grouped into ``stands'', each consisting of a linear-orthogonal pair of antennas, feedpoint-mounted electronics, a mast, and a ground screen as shown in Figure~\ref{fStand}.  Each antenna is a wire-grid ``bowtie'' about 3~m long, with arms bent downward at $45^{\circ}$ from the feedpoint in order to improve pattern uniformity over the sky.  The feedpoint is located 1.5~m above ground.  The ground screen is a 3~m $\times$ 3~m wire grid with spacing 10~cm $\times$ 10~cm and wire radius of about 1~mm. The primary purpose of the ground screen is to isolate the antenna from the earth ground, whose characteristics vary significantly as function of moisture content.  
%Zenith values of the effective aperture of a single dipole within a stand, including loss due to mismatch with the $100\Omega$ input impedance of the front end electronics, are estimated to be 0.25~m$^2$, 8.72~m$^2$, and 2.48~m$^2$ for 20~MHz, 38~MHz, and 74~MHz, respectively \cite{Ellingson11}.  
Zenith values of the effective aperture of a single dipole within a stand, multiplied by the loss due to mismatch with the $100\Omega$ input impedance of the front end electronics and accounting for radiation efficiency due to ground loss, are estimated to be 0.25~m$^2$, 8.72~m$^2$, and 2.48~m$^2$ for 20~MHz, 38~MHz, and 74~MHz, respectively \cite{Ellingson11}. 
Dividing out the impedance mismatch loss determined separately \cite{Hicks12}, the associated effective apertures are estimated to be 12.5~m$^2$, 19.0~m$^2$, and 5.4~m$^2$, respectively. 
However values for the actual {\it in situ} antennas are found to vary on the order of 25\% from antenna to antenna due to mutual coupling.  
Additional details of the design and analysis of these antennas can be found in \cite{Ellingson11} and \cite{Hicks12}. 

\begin{figure}
\begin{center}
%\vspace{3in}
%\psfig{file=figs/active_dipole_spectrum.eps,width=3.5in}
\psfig{file=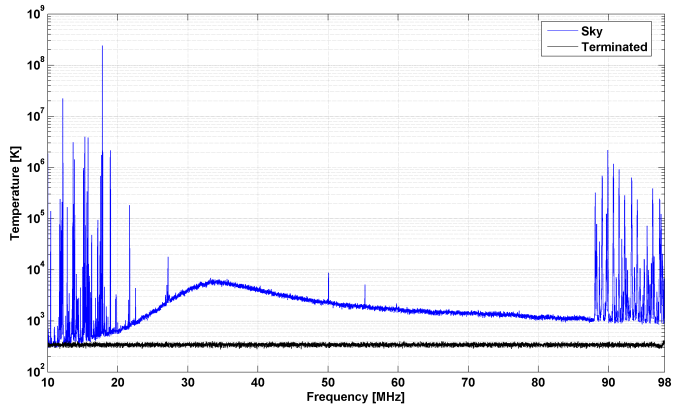,width=3.5in} %3.1
\end{center}
\caption{\label{fFEE}
Power spectral density 
%{\color{blue}(PSD)} 
measured by an LWA1 receiver, calibrated to the antenna terminals.  
{\it Top curve}: Result expressed as system temperature %(PSD/$k$)
(spikes are human-generated signals); 
{\it Bottom curve}: Same measurement made with a short circuit termination at the input, which provides an estimate of the internal contribution to the system temperature.  Spectral resolution: 6~kHz. Integration time: 10~s. Early afternoon local time.
}
\end{figure}

{\it Front end electronics.}  Each dipole is terminated into a pair of commercial InGaP HBT MMIC amplifiers (Mini-Circuits GALI-74) in a differential configuration, presenting a $100\Omega$ balanced load to the antenna.  This is followed by a passive balun which produces a $50\Omega$ single-ended signal suitable for transmission over coaxial cable, plus some additional gain to overcome cable loss.  The total gain, noise temperature, and input 1~dB compression point of the resulting front end electronics units are approximately 36~dB, 300~K, and $-18$~dBm respectively, and are approximately independent of frequency over 10--88~MHz \cite{Hicks12}.  
The gain and noise temperature of the feedpoint electronics are such that they dominate the noise temperature of the complete receiver chain, which is much less than the antenna temperature as is desirable (See Section~\ref{sBackground}).
The 1~dB compression point has been found to be satisfactory at the LWA1 site.  Although higher 1~dB compression would be better, this would be difficult to achieve without compromising noise temperature.  Additional information on design requirements for the front end electronics is available in \cite{Memo121}.
%
%Figure~\ref{fFEE} shows the measured power spectral density from the antenna (including impedance mismatch) and the front end electronics separately, confirming an internal noise temperature of about 300~K, 4:1 external noise dominance over 24--87~MHz,\footnote{It should be noted that the Galactic background-dominated antenna temperature varies diurnally over a range of about 35\% due to the rotation of the Earth (see \cite{ESP07} for an illustration); as a result the external noise dominance varies slightly over a 24-hour period.} and negligable level of intermodulation. 
The curve labeled ``Sky'' in Figure~\ref{fFEE} shows the noise temperature measured by an LWA1 receiver in normal operation (i.e., dipole attached), after calibration to account for the known gain of the electronics following the antenna terminals.  Thus, this is an estimate of the system temperature.  Also shown is the same measurement with the dipole terminals short-circuited, which is assumed to zero the noise delivered to the front end electronics unit without significantly changing its behavior, which is consistent with the findings of laboratory experiments.  These result confirm an internal noise temperature of about 300~K, 4:1 external noise dominance over 24--87~MHz,\footnote{It should be noted that the Galactic background-dominated antenna temperature varies diurnally over a range of about 35\% due to the rotation of the Earth (see \cite{ESP07} for an illustration); as a result the external noise dominance varies slightly over a 24-hour period.
} and negligable level of intermodulation.  
%{\color{blue}The noise temperature of the front end electronics have been independently measured in laboratory conditions and determined to be less than 273~K \cite{Hicks12}.}   
%

\begin{figure}
\begin{center}
%\vspace{3in}
\psfig{file=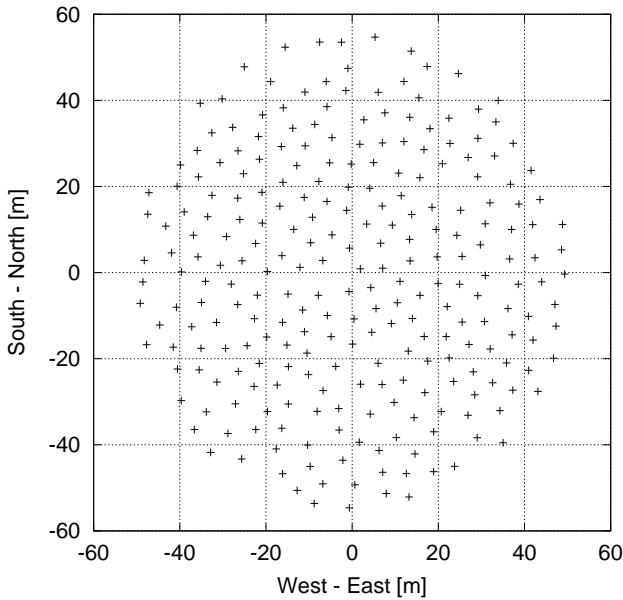,width=3.3in} %3.1
\end{center}
\caption{\label{fAG}
Arrangement of stands in the LWA1 array.  The minimum distance between any two masts is 5~m ($0.33\lambda$, $0.63\lambda$, and $1.23\lambda$ at 20~MHz, 38~MHz, and 74~MHz, respectively).  All dipoles are aligned North-South and East-West.  Outrigger stands are not shown.
%; $\phi=0$ is East.  %For additional information about the array geometry, see \cite{PIEEE_LWA}.
}
\end{figure}

{\it Array Geometry.}  LWA1 consists of 256 antenna stands (512 antennas) within a 100~m (East-West) $\times$ 110~m (North-South) elliptical footprint, plus two stands (4 antennas) which lie outside this footprint (the ``outriggers'').  The arrangement of stands is shown in Figure~\ref{fAG}.  The station diameter and number of stands per station were originally determined from an analysis of requirements for the LWA aperture synthesis imaging array, as detailed in \cite{PIEEE_LWA,K05}.  However, these choices are also appropriate for the present single-station instrument, as is demonstrated in this paper.  This choice of station aperture and number of stands results in a mean spacing between stands of about 5.4~m, which is $0.36\lambda$ and $1.44\lambda$ at 20~MHz and 80~MHz respectively.  To suppress aliasing, antennas are arranged in a pseudo-random fashion, with a minimum spacing constraint of 5~m in order to facilitate maintenance.  The elongation of the station aperture in the North-South direction improves main lobe symmetry for pointing towards lower declinations, including the Galactic Center, which transits at $Z\approx63^{\circ}$ as seen from the site.
%Traditional techniques for broadband array design require uniform spacings less than $0.5\lambda$ at the highest frequency of operation \cite{Hansen}.  This is for two reasons: (1) to prevent spatial aliasing, and (2) to use the strong electromagnetic coupling  to stabilize the scan impedance of the individual antennas, improving bandwidth.  However, to achieve this spacing at 80~MHz requires an increase in $N_a$ by more than a factor of 3, which is cost-prohibitive.  Implementing larger $N_a$ using a hierarchical (i.e., subarray-based) architecture allows more antennas at similar cost, but only by sacrificing the ability for beams to be steered independently over the entire sky.
%
%
% Area of a station = pi*(100m)^2/4 = 7854 m^2
% Area of station per stand = 30.7 m^2
% sqrt of above = 5.

{\it Cable System.} Connections between the front end electronics and the electronics shelter are by coaxial cables having lengths between 43~m and 149~m (excluding outriggers).  These cables have loss of about $-0.11$~dB/m at 88~MHz, and dispersive delays (that is, frequency-dependent delays in addition to the delay implied by the velocity factor) given by \cite{LWA170,RW53}:
\begin{equation}
\left(2.4~\mbox{ns}\right) \left( \frac{l}{100~\mbox{m}} \right) \left( \frac{\nu}{10~\mbox{MHz}} \right)^{-1/2} ~\mbox{,}
\end{equation}
where $l$ is length and $\nu$ is frequency.
Thus the signals arriving at the electronics shelter experience unequal delays, losses, and dispersion.  These can be corrected either ``in line'' in the station's digital processor (see below) or, for the all-dipoles modes, as a post-processing step.  In the current real-time beamforming implementation, the non-uniform dispersive delays are compensated for the center frequency of the highest-frequency tuning in a beam; thus there is some error over the bandwidth of the beam, and additional dispersion error for the lower-frequency tuning in the same beam.

\begin{figure}
\begin{center}
%\vspace{3in}
\psfig{file=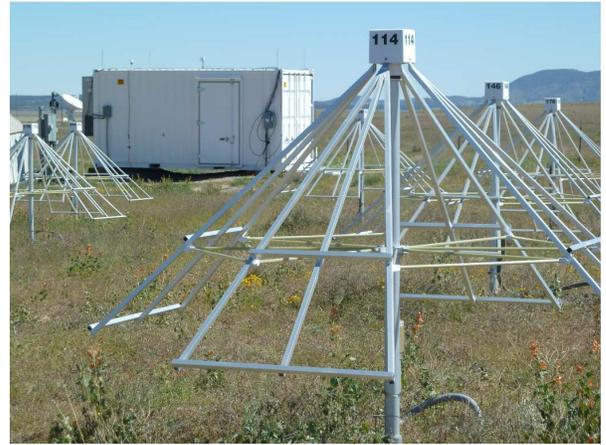,width=3.1in}
\end{center}
\caption{\label{fStandAndShelter}View from within the LWA1 array looking toward the equipment shelter.}
\end{figure}

{\it Receivers.} Upon arrival in the shelter (Figure~\ref{fStandAndShelter}), the signal from every antenna is processed by a direct-sampling receiver comprised of an analog section consisting of only gain and filtering, a 12-bit analog-to-digital converter (A/D) which samples 196 million samples per second (MSPS), and subsequent digital processing to form beams and tune within the digital passband.  
Digitization using fewer than 12 bits would be sufficient \cite{Memo121}, but the present design eliminates the need to implement gain control in the analog receivers and provides generous headroom to accommodate interference when it becomes anomalously large.
The choice of 196~MSPS ensures that strong radio frequency interference (RFI) from the 88--108 MHz FM broadcast band (see Figure~\ref{fFEE}) aliases onto itself, with no possibility of obscuring spectrum below 88~MHz.  
To accommodate the various uncertainties in the RFI environment, analog receivers can be electronically reconfigured between three modes:  A full-bandwidth (10--88~MHz) uniform-gain mode; a full-bandwidth dual-gain mode in which frequencies below about 35~MHz can be attenuated using a shelf filter\footnote{A ``shelf filter'' is a filter which has two adjacent passbands, with one passband (the “shelf”) having higher attenuation than the other.}; and a 28--54~MHz mode, which serves as a last line of defense should RFI above and/or below this range become persistently linearity-limiting.  In addition, the total gain in each mode can be adjusted over a 60~dB range in 2~dB steps, allowing fine adjustments to optimize the sensitivity-linearity tradeoff.  
%In practice, we use the dual-gain mode whenever observations below 35~MHz are not required, and the full bandwidth mode otherwise.  
Use of the 28--54~MHz mode has not been required to date.   

{\it Digital Beamforming.}   A detailed description of the LWA1 digital processor is provided in \cite{S+11} and is summarized here. Beams are formed using a time-domain delay-and-sum architecture. Delays are implemented in two stages: An integer-sample ``coarse'' delay is applied using a first-in first-out (FIFO) buffer operating on the A/D output samples, followed by a 28-tap finite impulse response (FIR) filter that implements an all-pass subsample delay. The filter coefficients can also be specified by the user, allowing the implementation of beams with custom shapes and nulls. The delay-processed signals are added to the signals from other antennas processed similarly to form beams. Four dual-polarization beams are constructed in this fashion, each fully-independent and capable of pointing anywhere in the sky. Each beam is subsequently converted to two independent ``tunings'' of up to 16~MHz bandwidth (4-bits ``I'' + 4-bits ``Q'' up to 19.6~MSPS) each, with each tuning having a center frequency independently-selectable from the range 10--88 MHz. 
Both polarizations and both tunings of a beam emerge as a single stream of user datagram protocol (UDP) packets on 10~Gb/s ethernet. 
Thus there are four ethernet output cables, with each one representing two center frequencies from a particular pointing on the sky.
The maximum data rate (ignoring protocol bits) on each ethernet cable carrying beam data is therefore 19.6 MSPS $\times$ 8~bits/sample $\times$ 2 polarizations $\times$ 2 tunings = 627.2~Mb/s.

%--- architectural pictures of BF/DRX, TBW, TBN

{\it All-Sky Modes.}  Simultaneously with beamforming, LWA1 is able to coherently capture and record the output of all its A/Ds, where each A/D corresponds to one antenna. This can be done in two distinct modes. The ``transient buffer -- wideband'' (TBW) mode allows the raw ($\approx78$~MHz) 12-bit output of the A/Ds to be collected in bursts of 61 ms at a time, and $\approx5$ minutes is required to write out the captured samples. The ``transient buffer -- narrowband'' (TBN) mode, in contrast, allows a single tuning of $\approx70$~kHz bandwidth to be recorded continuously for up to 20~hours.
%, resulting in an output data rate of $\sim113$ MB/s for the array (which is the maximum that a data recorder can absorb, thus setting the maximum TBN bandwidth).
These two modes share the same 10 Gb/s ethernet output from the digital processor, and thus are mutually exclusive. However, the TBW/TBN output is distinct from the four beam outputs and runs simultaneously with all four beams.

{\it Data Recorders/Spectrometers.} The limited data rate of the internet connection from the LWA1 site makes data transfer from the site impractical for observations longer than a few minutes.  Instead, each beam output and the TBW/TBN output is connected to a dedicated data recorder (DR). A DR is a computer that records the UDP packets to a ``DR storage unit'' (DRSU). 
Currently, a DRSU consists of five 2~TB drives (10 TB total) in a 1U rack-mountable chassis, configured as an eSATA disk array. 
Each DR can host 2 DRSUs. At the maximum beam bandwidth, each DRSU has a capacity of about $30$ hours of observation.  The data saved is fully coherent and Nyquist-sampled.  
Alternatively, the DRs can be used in ``spectrometer mode'' in which they continuously compute 32-channel\footnote{Increased to 1024 channels since this paper was originally submitted.} fast Fourier transforms (FFTs) on the incoming beam data independently for each beam and tuning, and then time-average the FFT output bins to the desired time resolution.  
This results in a dramatic reduction in data volume, but is not suitable for all observing projects.

{\it PASI.} The "Prototype All-Sky Imager" (PASI) is a software-defined correlator/imager currently operational at LWA1 using the TBN data stream.  
It consists of a cluster of 4 server-class computers with Nehalem multicore processors interconnected by an Infiniband switch.  
PASI images nearly the whole sky in the Stokes $I$ and $V$ parameters many times per minute, continuously and in real time, with an average duty cycle of better than 95\%.  (The $V$ parameter is useful for discriminating circularly-polarized emission such as that from Jupiter.)
PASI does this by cross-correlating the dipole data streams, producing a sampling of ``visibilities'' within the station aperture.  
These visibilities are then transformed into sky images using the NRAO's Common Astronomy Software Applications (CASA) software library.\footnote{http://casa.nrao.edu/}  
Figures~\ref{fPASI1} and \ref{fPASI2} were obtained from PASI.

%========================================================
\section{\label{sCal}Array Calibration}
%========================================================

In this section we describe the technique currently in use to calibrate the array for real-time beamforming.  Since LWA1 uses delay-and-sum beamforming, the problem is to determine the set of delays which, when applied to the dipole (A/D) outputs, results in a beam with maximum directivity in the desired direction, subject to no other constraints. (In \cite{Ellingson11}, this is referred to as ``simple'' beamforming.  Other approaches are possible but are not considered in this paper.)  In principle these delays can be estimated {\it a priori}, since the relevant design values (in particular, cable lengths) are known in advance.  In practice this does not work well due to errors in presumed cable lengths and the accumulation of smaller errors associated with distribution of signals within equipment racks.  Thus, delays must be measured using external stimulus signal(s) while the instrument is in operation.  
 
The approach used here is to decompose the problem into a set of narrowband calibration problems which are solved using data collected using LWA1's TBN mode, and then to extract delays by fitting a model of the presumed cable response (including dispersion) to the measured phase responses.  
The narrowband decomposition is justified by the fact that the maximum time-of-flight between any two antennas in the station array is $\le367$~ns, which is much less than the TBN inverse bandwidth (70~kHz)$^{-1} \approx 14~\mu$s.  As described below, the outriggers play an important role in the calibration; the maximum time-of-flight between any antenna in the station array and the outrigger (maximum separation $\approx390$~m) is $\le1.3~\mu$s.

\begin{figure}
\begin{center}
%\vspace{3in}
\psfig{file=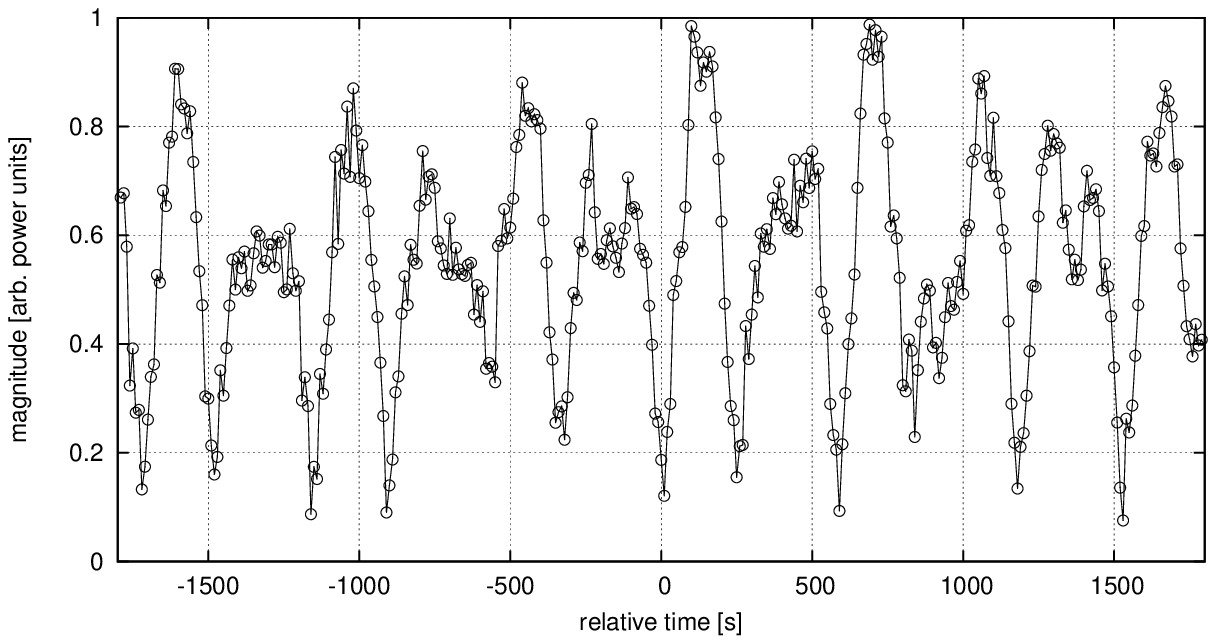,width=3.3in} %3.1
\psfig{file=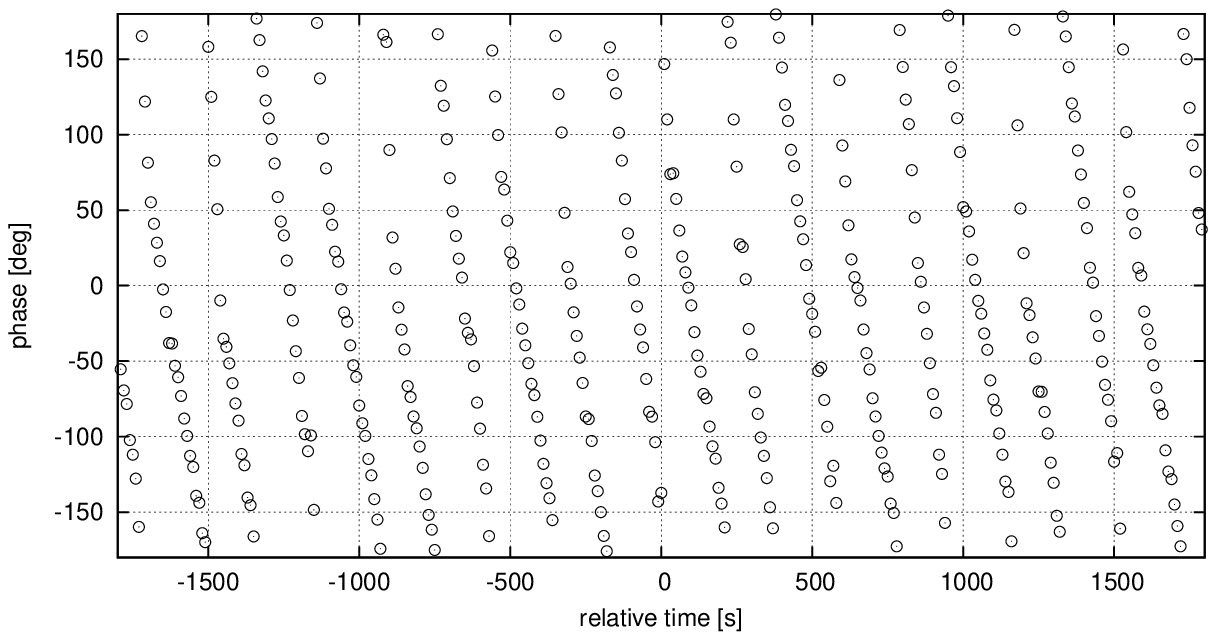,width=3.3in} %3.1
\end{center}
\caption{\label{fFringes} Fringes between a dipole on the far West side of the array and an outrigger dipole located 389~m to the East. {\it Top:} Magnitude. {\it Bottom:} Phase. Each point represents 10~s integration. Start time is 18:50 LST.}
\end{figure}

\begin{figure}
\begin{center}
%\vspace{3in}
\psfig{file=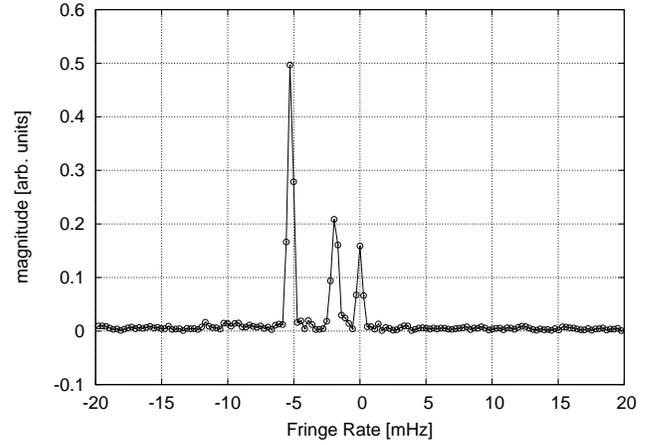,width=3.4in} %3.1
\end{center}
\caption{\label{fFringeTransform} Fringe rate spectrum corresponding to Figure~\ref{fFringes}.  This result is dominated by two bright sources (these happen to be the radio galaxy Cygnus~A (Cyg~A) and the supernova remnant Cassiopeia~A (Cas~A)) and a ``DC'' term. The DC term is the ``all sky'' contribution predicted in \cite{Ellingson11}, and stands as additional evidence that the observation is strongly external noise-dominated. Many other sources are present, but are orders of magnitude weaker. The frequency resolution is 279~$\mu$Hz.}
\end{figure}

The narrowband procedure is described in rigorous mathematical detail in \cite{LWA184}, and summarized here. Each narrowband calibration relies on the ability to identify the response due to a point source in the array output.  This is confounded by the problem that there are typically multiple bright sources present in the sky, and (as explained in Section~\ref{sBackground}) the problem that Galactic noise dominates the system temperature and thus appears in the data as a bright, non-uniformly distributed source.  To suppress the effect of distributed features (including the Galactic center and Galactic plane) as well as discrete sources other than the source of interest, we use a ``fringe rate filtering'' technique.  Fringe rate filtering is essentially a simple narrowband version of the delay / delay-rate filtering technique of Parsons \& Backer (2009) \cite{PB09}.  In this technique, each dipole in the array is correlated with a corresponding outrigger dipole.  The electrically-large spacing makes the correlation relatively insensitive to spatial structure in the sky noise intensity having large angular scales (see \cite{Ellingson11} for examples.)  The contributions to the correlation due to individual discrete sources exhibit time-varying phase due to the apparent rotation of the sky; in astronomical parlance, these source-dominated correlations are referred to as ``fringes''; see Figure~\ref{fFringes} for an example.  The rate of phase rotation in the fringes depends on the positions of the antennas being correlated relative to the direction to the sources.  A time-to-frequency Fourier transform of the correlation over an interval greater than the reciprocal of the smallest source-specific fringe rate yields a ``fringe rate spectrum'', in which discrete sources are apparent as localized components; an example is shown in Figure~\ref{fFringeTransform}.  ``Fringe rate filtering'' refers to the process of selecting just one of these components, suppressing all others (which can be done by matched filtering (``fringe stopping'') in the time domain or by excision in the fringe rate domain); the resulting correlation then represents the response of just one source.    This of course presumes that sources are sufficiently separated in fringe rate: This is one reason for the use of outriggers, since fringe rate is proportional to antenna separation.  Further improvement is possible by proper scheduling of the associated observations; that is, choosing times in which suitable sources have sufficiently different fringe rates.  For completeness, the positions and strengths of the sources used in the example shown in Figures~\ref{fFringes} and \ref{fFringeTransform} are given in Table~\ref{t1}.  
\begin{table}[h]
\begin{center}
\begin{tabular}{llllll}
\hline
%Source & RA (h:m:s)  & Dec (d:m:s)   & Az (d:m:s) & $Z$ (d:m:s) & Flux density \\
Source & RA  & Dec   & Az & $Z$ & Flux density \\
\hline
\hline
%Cyg A  & 19:59:28.35 & $+$40:44:02.1 & 46:07:35   & 10:11:14    & 17.06~kJy \\ %& TrnTm 23:21 TrnAlt 83:19
%Cas A  & 23:23:26.00 & $+$58:48:00.0 & 38:29:11   & 46:26:32    & 17.13~kJy \\ %& TrnTm 02:49 TrnAlt 65:13 
Cyg A  & 19h 59$'$ & $+$40$^{\circ}$44$'$ & 46$^{\circ}$08$'$  & 10$^{\circ}$11$'$ & 17.06~kJy \\ %& TrnTm 23:21 TrnAlt 83:19
Cas A  & 23h 23$'$ & $+$58$^{\circ}$48$'$ & 38$^{\circ}$29$'$  & 46$^{\circ}$27$'$ & 17.13~kJy \\ %& TrnTm 02:49 TrnAlt 65:13 
\hline
\end{tabular}
\end{center}
\begin{tiny}
Right ascension (RA) and declination (Dec) are given for the J2000 epoch.\\  
Azimuth (Az) and zenith angle ($Z$) are indicated for the midpoint of the observation.\\  
The flux densities are the 73.8~MHz values given in LWA Memo 155 \cite{LWA155}, scaled to 74.03~MHz using spectral indices of $-0.58$ and $-0.72$ for Cyg~A and Cas~A respectively, and Cas~A is further scaled by $-0.75$\%/yr \cite{HK}.
\end{tiny}
\caption{\label{t1} Astronomical source data pertaining to Figures~\ref{fFringes} and \ref{fFringeTransform}.}
\end{table}
%     MDT Aug 19, 2011 22:42:50 : Cyg A : RA 19:59:28.35 DEC 40:44:02.1 Az  46:07:35 Alt 79:48:46 Z 10:11:14 TrnTm 23:21 TrnAlt 83:19
%                                 Cas A : RA 23:23:26.00 DEC 58:48:00.0 Az  38:29:11 Alt 44:33:28 Z 46:26:32 TrnTm 02:49 TrnAlt 65:13 
%At the mid-point of this observation (22:42 MDT), the positions of Cyg~A and Cas~A were as indicated in Table~\ref{t1}.

This procedure is repeated for every antenna in the array, yielding a set of complex-valued coefficients associated with one direction at one frequency.  One then iterates over frequency.  There is currently no iteration over direction; that is, only one direction is considered.  Different directions give different coefficients, for two reasons:  First, because the geometrical delays are different; however these are easily equalized.  The second reason is because the antenna patterns are made unequal by the effects of mutual coupling.  The effects of mutual coupling on the antenna patterns is difficult to know precisely.  However since our goal is to compute (broadband) delays, as opposed to (narrowband) phases and magnitudes, calibration in a single direction is sufficient, as will be demonstrated in the next section.  
The entire procedure requires on the order of a week, typically observing over the same one-hour LST range at a different frequency each day. 

%========================================================
\section{\label{sBeam}Beamforming Performance}
%========================================================

%We now show examples demonstrating the performance of LWA1 beamforming.  

\subsection{\label{s67}Multi-frequency Drift Scan at $Z=6.7^{\circ}$}

\begin{figure}
\begin{center}
%\vspace{3in}
%\psfig{file=figs/CygA_drift.eps,width=3.1in}
\psfig{file=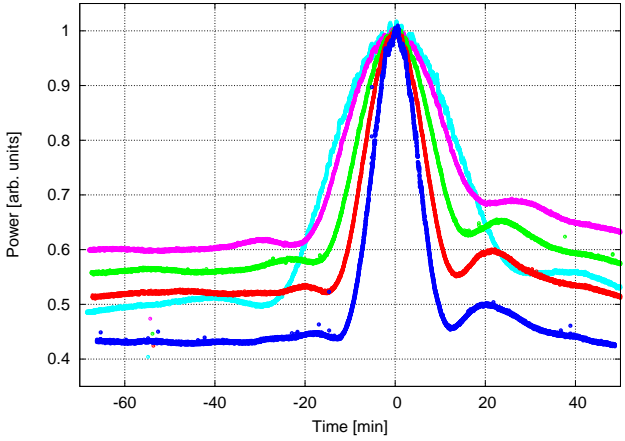,width=3.4in} %3.1
\end{center}
\caption{\label{fCygA_drift} Simultaneous drift scans of Cyg~A at 85.00, 74.03, 62.90, 52.00, and 37.90~MHz. Scans can be identified by peak width, which increases with decreasing frequency. All scans are normalized to a common maximum.  Bandwidth: 211~kHz, Integration time: 786~ms, Single (N-S) polarization, no RFI mitigation.}
\end{figure}

\begin{figure}
\begin{center}
%\vspace{3in}
\psfig{file=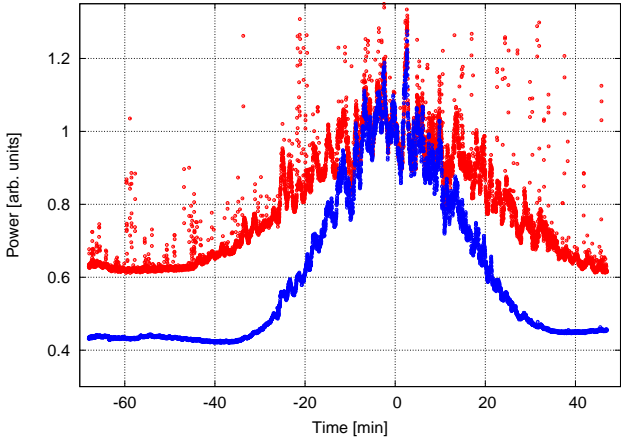,width=3.4in} %3.1
\end{center}
\caption{\label{fCygA_drift_lf} Simultaneous drift scans of Cyg~A at 28.80 and 20.50~MHz. Scan magnitudes are normalized.  Bandwidth: 211~kHz, Integration time: 786~ms, Single (N-S) polarization, no RFI mitigation.  }
\end{figure}

Figures~\ref{fCygA_drift} and \ref{fCygA_drift_lf} show a single transit drift scan of Cyg~A; i.e., the output of a fixed beam pointed at the location in the sky at which Cyg~A achieves upper culmination: $Z=6.7^{\circ}$, $0^{\circ}$ north azimuth.   
Cyg~A is useful as a test source both because it is the strongest time-invariant and unresolved astronomical source visible from the LWA1 site, and also because it transits close to the zenith.  
The origin in both plots corresponds to the time at which Cyg~A is expected to peak; Figure~\ref{fPASI1} shows the sky at this time.
The results for a narrow bandwidth at 8 distinct center frequencies ranging from 20.5 through 85.6~MHz are shown.
The large peak in each curve is Cyg~A; the small peak to the right in some curves is the Galactic plane passing through the beam a short time later (apparent only at higher frequencies and unresolved in lower-frequency curves). 
Note that Cyg~A is clearly detectable at all frequencies, demonstrating the very wide instantaneous bandwidth of the instrument.  
Also apparent is the broadening of the beamwidth with decreasing frequency, as expected.   
The lowest two frequencies (28.80 and 20.50~MHz) are shown in a separate figure for clarity; both frequencies are apparently strongly affected by ionospheric scintillation, and 20.05~MHz is strongly affected by RFI.
 
The drift scans for each frequency in Figure~\ref{fCygA_drift} and \ref{fCygA_drift_lf} are normalized to the same peak value to facilitate comparison of beam width.
Using the method described in Appendix~\ref{sBeamwidth}, we find full-width at half-maximum (FWHM) beamwidths of $2.8^{\circ}$, $3.5^{\circ}$, $4.6^{\circ}$, $5.8^{\circ}$, $7.5^{\circ}$, $8.0^{\circ}$ and $10.9^{\circ}$ for 85.00, 74.03, 62.90, 52.00, 37.90, 28.80, and 20.50~MHz respectively.  
Note that these measurements are in planes of constant declination; i.e., not planes of constant azimuth nor constant elevation.
%
%{\color{blue}
%For comparison, values for planes of constant azimuth can be estimated using the following crude model:}
%%
%\begin{equation}
%\mbox{FWHM} \approx 2.2^{\circ} \times \left( \frac{\mbox{74~MHz}}{\nu} \right) \sec^2{Z}~\mbox{,}
%\label{eFWHM}
%\end{equation}
%
%where $2.2^{\circ}$ is the expected zenith FWHM of a uniformly-illuminated aperture having the same dimensions as LWA1, at 74~MHz  \cite{PIEEE_LWA}. 
%The measurements are roughly 20--60\% larger than predicted by the above model.  This is due in part to the proximity of Cyg~A to the Galactic plane, and the resulting ``confusion'' of the Cyg A emission with diffuse Galactic emission filling the beam.  This is evident by comparison with the same measurement performed using the source 3C123, which lies further from the Galactic plane but transits at a similar zenith angle ($Z=4.4^{\circ}$).  For 3C123 we obtain FWHM = $2.7^{\circ}$ at 74.03~MHz, which is 23\% less than the Cyg~A value.  Due to this limitation in the measurement technique, the FWHM values reported here should be considered upper bounds as opposed to estimates of the actual values. 
%
The expected zenith FWHM of a uniformly-illuminated circular aperture having the same minimum dimensions as LWA1 is $2.3^{\circ}$ at 74~MHz; thus the measurements are roughly 20--60\% larger than predicted by the above model.  This is due in part to the proximity of Cyg~A to the Galactic plane, and the resulting ``confusion'' of the Cyg A emission with diffuse Galactic emission filling the beam.  This is evident by comparison with the same measurement performed using the source 3C123, which lies further from the Galactic plane but transits at a similar zenith angle ($Z=4.4^{\circ}$).  For 3C123 we obtain FWHM = $2.7^{\circ}$ at 74.03~MHz, which is quite close to the expected value.  Due to this limitation in the measurement technique, the FWHM values reported here should be considered upper bounds as opposed to estimates of the actual values. 

We now consider beam sensitivity.  Because LWA1 is strongly Galactic noise-limited, beam directivity is not a reliable metric of sensitivity.  Instead, we consider sensitivity in terms of system equivalent flux density (SEFD), which is defined as the strength of an unresolved (point) source that doubles the power at the output of the beam relative to the value in the absence of the source.  Because the flux density of Cyg~A is known (Table~\ref{t1}), beam SEFD can be determined directly from drift scans using the method described in Appendix~\ref{sSEFD}.
This method yields SEFD $= 16.1$~kJy for 74.03~MHz.  To obtain the source flux $S$ at some other frequency $\nu$, we apply the known spectral index $-0.58$ for Cyg~A \cite{Baars77}: 
\begin{equation}
S({\nu}) = S({\nu_0}) \left( \frac{\nu}{\nu_0} \right)^{-0.58}~\mbox{,}
\end{equation}
where $\nu_0$ is a frequency at which the flux is already known.
From this we estimate SEFD = 11.0, 21.5, 28.8, 21.5, 18.2, and 47.0~kJy for 85.00, 62.90, 52.00, 37.90, 28.80, and 20.50~MHz respectively.  Due to the variables described in Section~\ref{sBackground} it is difficult to know even within 10's of percent what values to expect; however it is shown in the next section that these results are consistent with the results of simulations in previous work.  It should also be emphasized that the flux density of radio sources (including Cyg A and sources discussed in the next section) is known only approximately below 74 MHz, and that use of the 74 MHz flux density and spectral index to calculate flux density at lower frequencies is another source of error.

\subsection{Sensitivity \& Beamwidth vs.\ Frequency \& Elevation}

\begin{figure}
\begin{center}
%\vspace{1.1in} {\it [Under construction.]} \vspace{1.1in}
\psfig{file=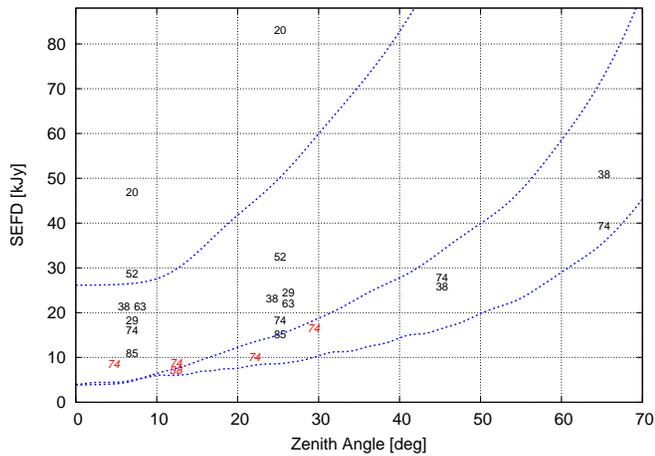,width=3.5in} %3.1
\end{center}
\caption{\label{fSEFD} Sensitivity (SEFD) vs.\ $Z$ obtained from drift scans.  The numbers used as data markers indicate the frequency rounded to the nearest MHz.  Markers in red italic font represent transit drift scans of the sources 3C123 ($Z=4.4^{\circ}$), Tau~A ($12.1^{\circ}$), Vir~A ($21.8^{\circ}$), and 3C348 ($29.1^{\circ}$); all others are Cyg~A.  The curves are predictions from Fig.~8 of \cite{Ellingson11} for (bottom to top) 74, 38, and 20~MHz.  The use of both polarizations is assumed.}
\end{figure}

The same procedures described in the previous section have been used to determine sensitivity at lower elevations using Cyg~A and the strong sources Tau~A, Vir~A, 3C123, and 3C348.  The results are summarized in Figure~\ref{fSEFD}.  Also shown in this figure are predictions originally shown in Fig.~8 of \cite{Ellingson11}.
Note that SEFD depends both on pointing with respect to celestial coordinates as well as sidereal time (for the reasons explained in Section~\ref{sBackground}); whereas the predictions of \cite{Ellingson11} assume a uniformly bright sky.
%Further, the simulations assume standard constitutive parameters parameters for regional conditions, whereas the actual constitutive parameters are significantly affected by soil moisture and other local factors.
Thus, precise agreement cannot be expected.  
Nevertheless, the levels and trends with $Z$ and frequency appear to be consistent with the predictions. 
We do note however that the agreement for 38~MHz appears to be somewhat worse than the agreement at other frequencies.  We do not currently have an explanation for this.

\begin{figure}
\begin{center}
%\vspace{1.1in} {\it [Under construction.]} \vspace{1.1in}
\psfig{file=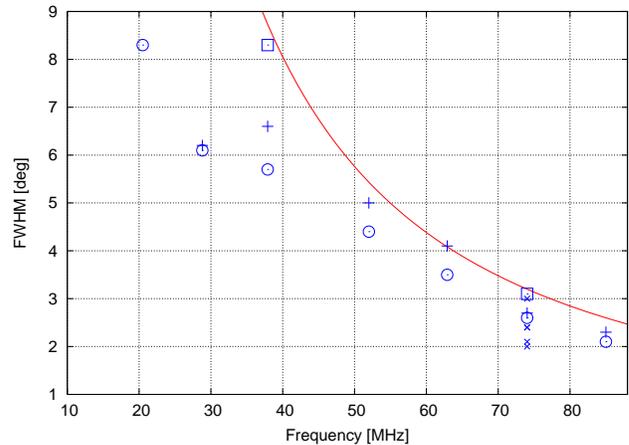,width=3.4in} %3.1
\end{center}
\caption{
\label{fFWHM} 
Beamwidth (FWHM) vs.\ frequency obtained from drift scans. 
Markers indicate $Z$: ``$\circ$'',  $6.7^{ \circ }$; ``$+$'', $25^{\circ}$; square, $45^{\circ}$; all using Cyg~A. ``$\times$'' are 74.03~MHz results using transit drift scans from the same additional sources identified in the caption of Figure~\ref{fSEFD}, all at 74.03~MHz.   
The solid line is the upper bound described in the text.
}
\end{figure}

Figure~\ref{fFWHM} shows a summary of FWHM measurements from drift scans.  Beamwidth is difficult to measure at large $Z$ because the drift scan peaks become simultaneously broad and weak; thus the data shown is limited to $Z \le 45^{\circ}$.   
%Also shown are the expectations based on the model (Equation~\ref{eFWHM}).  
Note that Cyg~A measurements generally indicate FWHM higher than expected, whereas measurements using other sources are closer to expectations; again this is due in some part to the inability of the method of Appendix~\ref{sBeamwidth} to account for the excess emission associated with the nearby Galactic plane.  
Also shown in Figure~\ref{fFWHM} is an empirical upper bound based on all measurements (regardless of $Z$) considered to date:
\begin{equation}
\mbox{FWHM} < 3.2^{\circ} \times \left( \frac{\mbox{74~MHz}}{\nu} \right)^{1.5}~\mbox{.}
\end{equation}
We intend to improve on this bound in future measurements by using interferometric methods, as discussed in Section~\ref{sConc}.

%========================================================
\section{\label{sConc}Conclusions}
%========================================================

This paper described the design of the LWA1 radio telescope, array calibration technique, and the results of commissioning experiments to confirm beamforming performance.  
The ``single point'' delay calibration technique currently in use at LWA1 was described in Section~\ref{sCal}.  In this technique, the delays are estimated from phase estimates over a range of frequencies using narrowband ``all-dipoles'' (TBN) observations of the sky.  This technique uses a fringe rate filtering technique which is robust to the presence of multiple bright sources including both unresolved and spatially-extended sources.  
Real time delay-and-sum beamforming was demonstrated in Section~\ref{sBeam}.  
Sensitivity findings are summarized in Figure~\ref{fSEFD}, and were shown to be consistent with predictions made in \cite{Ellingson11}.  
Beamwidth findings are summarized in Figure~\ref{fFWHM}. Due to the limitations of the drift scan measurement technique employed in this paper, we can currently only upper-bound our estimates of beamwidth.  In future measurements we intend to cross-correlate drift scan beams with outrigger dipoles, which is expected to suppress the contribution of diffuse Galactic emission and result in more accurate measurement of beamwidth.  The same technique will enable characterization of sidelobes, which is not possible with the drift scan technique.  We also plan to conduct measurements to characterize polarization performance.  The results presented here combined with early science results described in \cite{LWAFL} confirm that LWA1 performance is consistent with expectations and that the instrument is ready for science operations.

%%%%%%%%%%%%%%%%%%%%%%%%%%%%%%%%%%%%%%%%%%%%%%%%
%%%%%%%%%%%%%%%%%%%%%%%%%%%%%%%%%%%%%%%%%%%%%%%%  
\useRomanappendicesfalse
\appendices
%%%%%%%%%%%%%%%%%%%%%%%%%%%%%%%%%%%%%%%%%%%%%%%%
%%%%%%%%%%%%%%%%%%%%%%%%%%%%%%%%%%%%%%%%%%%%%%%% 

%========================================================
\section{\label{sBeamwidth}Estimation of Beamwidth from a Drift Scan}
%========================================================

The peak-to-baseline ratios achieved in drift scans such as those shown in Figures~\ref{fCygA_drift} and \ref{fCygA_drift_lf} are too low to facilitate calculation of beam FWHM by direct inspection.  However it is possible to estimate FWHM by modeling the drift scan as the sum of a Gaussian function representing the power pattern of the main lobe plus a constant noise floor, solving for the coefficient of time in the exponent of the Gaussian function, and then calculating the associated width of the Gaussian function alone.  In this appendix we first derive the result, and then we justify the use of the Gaussian approximation.

The drift scan is modeled as:
\begin{equation}
y(t) = A + B e^{-\gamma t^2}
\end{equation}
where $A$ is the magnitude of the constant noise baseline, $B$ is the peak magnitude of the Gaussian function modeling the main lobe, and $t$ is time.  Evaluating this expression at the time $t=0$, taken to be the time at which the peak occurs, and time $t_1$, a short time later, we have:
\begin{equation}
y(0) = A + B
\end{equation}
\begin{equation}
y(t_1) = A + B e^{-\gamma t_1^2}~\mbox{.}
\end{equation}
Solving for $\gamma$:
\begin{equation}
\gamma = \frac{\ln{(y(0)-A)}-\ln{(y(t_1)-A)}}{t_1^2}~\mbox{.}
\label{eGAa}
\end{equation}
FWHM corresponds to the time interval between the half-maximum points of the Gaussian function, which is easily found to be
\begin{equation}
\mbox{FWHM} = 2\sqrt{\frac{\ln{2}}{\gamma}}~\mbox{.}
\label{eGAf}
\end{equation}

\begin{figure}
\begin{center}
\psfig{file=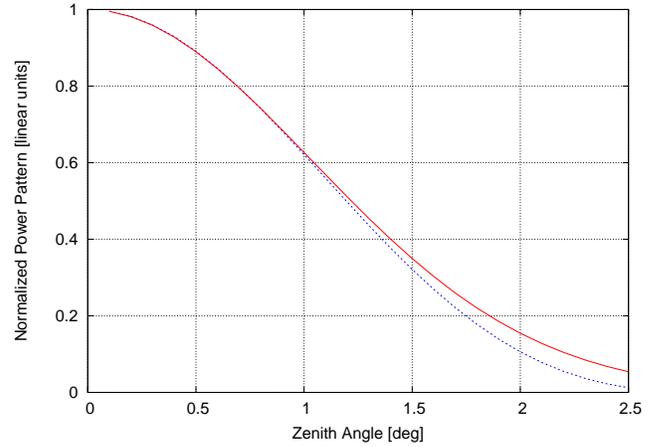,width=3.4in} %3.1
\end{center}
\caption{\label{fML} Comparison of the Gaussian main lobe model $y(cZ)$ ({\it blue/dashed}) to the main lobe power pattern $F^2(Z)$ ({\it red/solid}) for $a/\lambda=12.3$
%{\color{blue}\sout{, modeling an LWA1 zenith-pointing beam at 74~MHz}}
.}
\end{figure}

%{\color{blue} We now show that the Gaussian model is a reasonable approximation to the exact result for a uniformly-weighted circular aperture.  Note the LWA1 aperture is not exactly circular, nor is it exactly uniformly-weighted (due to variations in antenna responses due to mutual coupling); nevertheless, this is reasonable approximation given the likely larger errors inherent in FWHM measurement by drift scan measurement as described in the text, and also because a more accurate model is not readily available.}
We now show that the Gaussian model is a reasonable approximation to the exact result for the zenith-pointed beam due to a uniformly-weighted circular aperture.  Note the LWA1 aperture is not exactly circular, nor is it exactly uniformly-weighted (due to variations in antenna responses due to mutual coupling); our purpose here is simply to demonstrate that the Gaussian model reasonably represents the kind of beam that we expect to see from this type of antenna system.
The normalized (i.e., maximum magnitude $= 1$) pattern function of a uniformly-weighted circular aperture lying in the plane of the ground is \cite{ST}:
\begin{equation}
F(Z) = 2 \frac{ J_1\left( \beta a \sin{Z} \right) }{ \beta a \sin{Z} }~\mbox{,}
\label{ePP}
\end{equation}
where $a$ is the radius of the aperture, $\beta = 2\pi/\lambda$, and $\lambda$ is wavelength.  Thus the normalized power pattern is $F^2(Z)$.  Let us assume that the main lobe of $F^2(Z)$ can be modeled as a Gaussian function $y(Z/c)$, where $c$ is the angular rate of drift.  For purposes of demonstration let us assume $a/\lambda=12.3$ (100~m diameter at 74~MHz).  We take $A=0$, $y(0)=1$, $y(t_1=Z_1/c)=F^2(Z_1)=0.9$, and then use Equation~\ref{ePP} to find $ct_1=0.475^{\circ}$.  Figure~\ref{fML} compares $F^2(Z)$ to the derived model $y(Z/c)$ with $\gamma$ obtained from Equation~\ref{eGAa}.  Note that the model error at the half-maximum point is very small, justifying the use of the Gaussian model for this purpose.  

Finally, we note that the model has not been demonstrated to be appropriate for beam pointings far from the zenith.  However, we note that we do not have a ``true'' model available for this case either, since the patterns of individual antennas combined with mutual coupling make this difficult to know. This caveat should be considered when evaluating results computed for low-elevation pointings.  

%========================================================
\section{\label{sSEFD}Calculation of SEFD from Drift Scans}
%======================================================== 

When the flux density of the source is known, beam system equivalent flux density (SEFD) can be estimated from drift scans such as those shown in Figures~\ref{fCygA_drift} and \ref{fCygA_drift_lf}.  A derivation follows.  The power density $P_s$ captured in one polarization of the beam associated with an unpolarized source having flux density $S$ is given by 
\begin{equation}
P_s = \frac{1}{2} S A_e
\end{equation}  
where $A_e$ is the effective aperture.  Similarly, the power density $N$ captured in one polarization of the beam associated with the total available noise is given by
\begin{equation}
N = \frac{1}{2} k T_{sys}   
\end{equation}  
where $k$ is Boltzmann's constant and $T_{sys}$ is the system noise temperature, assuming this to be external noise-dominated.  Thus the signal-to-noise ratio at the beam output is
\begin{equation}
\frac{P_s}{N} = \frac{S A_e}{k T_{sys}}~\mbox{.}
\end{equation}  
In these terms, the SEFD for this polarization is defined as the value of $S$ which results in $P_s/N=1$; thus:
\begin{equation}
\mbox{SEFD} = \frac{k T_{sys}}{A_e}~\mbox{.}
\end{equation}  
Let $P_1 = P_s + N$ be the power density measured at the peak of the drift scan, and let $P_0 = N$ be the power density in the absence of the source.  Note $P_0$ can be estimated from the approximately-constant noise baseline on either side of the peak.  Using the above definitions, we find
\begin{equation}
\frac{P_1}{P_0} = \frac{\frac{1}{2}S A_e + \frac{1}{2} k T_{sys} }{ \frac{1}{2} k T_{sys} } = \frac{S}{\mbox{SEFD}}+1~\mbox{.}
\end{equation}  
Solving for SEFD:
\begin{equation}
\mbox{SEFD} = S\left( \frac{P_1}{P_0} -1 \right)^{-1}~\mbox{.}
\end{equation}
Thus, given the source flux density, the SEFD can be obtained from $P_1$ and $P_0$, which themselves can be read from a single-polarization drift scan.  
%Assuming an unpolarized source and orthogonal polarizations, the two-polarization SEFD is lower by a factor of $\sqrt{2}$.  
%This is not exact, since the patterns of the two polarizations are slightly different; however the corresponding error is typically small, as is confirmed by comparison of the drift scans for polarization pairs.

%====================================================================
%====================================================================
% use section* for acknowledgement

\section*{Acknowledgments}
% optional entry into table of contents (if used)
%\addcontentsline{toc}{section}{Acknowledgment}

The authors acknowledge contributions to the design and commissioning of LWA1 made by  
S.\ Burns 
of Burns Industries; 
%{\it (names)}
%of Los Alamos National Laboratory;
L.\ D'Addario,
R.\ Navarro,
M.\ Soriano,
E.\ Sigman, and
D.\ Wang
of NASA Jet Propulsion Laboratory;
N.\ Paravastu 
and
H.\ Schmitt
of the U.S. Naval Research Laboratory; 
S.\ Tremblay 
of the University of New Mexico; and
M.\ Harun,
Q.\ Liu, and
C.\ Patterson
of Virginia Tech.
The authors acknowledge the helpful comments of M. Davis.

Construction of LWA1 has been supported by the U.S.\ Office of Naval Research under Contract N00014-07-C-0147. Support for operations and continuing development of LWA1 is provided by the National Science Foundation under Grant AST-1139963.  Basic research in radio astronomy at the Naval Research Laboratory is supported by 6.1 base funding. The authors acknowledge the support of the National Radio Astronomy Observatory.

\end{document}